\documentclass[11pt]{article}

\usepackage[margin=1in]{geometry}
\usepackage{amsmath,amssymb,amsthm}
\usepackage{mathtools,bm}
\usepackage{graphicx}
\usepackage{natbib}
\usepackage{url}
\usepackage{algorithm2e}
\usepackage{microtype}
\usepackage{enumitem}
\usepackage{booktabs}
\usepackage{xcolor}
\usepackage[hidelinks]{hyperref}

\newcommand{\E}{\mathbb E}

\usepackage{xcolor}

\newtheorem{theorem}{Theorem}
\newtheorem{lemma}[theorem]{Lemma}

\theoremstyle{definition}

\theoremstyle{remark}

\title{Learning under Bounded Manipulation\\
for Quadratic Incentive Design}

\author{%
Arda G\"u\c{c}l\"u
\quad Subhonmesh Bose
\quad Tamer Ba\c{s}ar\\[0.5em]
{\small Coordinated Science Laboratory}\\
{\small University of Illinois Urbana-Champaign, USA}\\[0.3em]
}

\date{} 

\begin{document}
\maketitle

\begin{abstract}
We study a repeated quadratic incentive design game in which a leader learns a follower's private type from noisy observations of the follower's actions and uses her estimates in an affine incentive rule. The follower knows the leader's learning policy, which incorporates past observations to produce estimates of the follower's type, and can deviate from his myopic best response to influence the leader's future estimates. We ask how the follower's ability to manipulate these observations affects the leader's performance. Under a manipulation budget $C_T$ constraining the expected cumulative squared deviation of the follower's actions from his myopic best response, we establish a minimax regret rate of order $\max\{\log T, \min\{C_T,T\}\}$ for approximately rational followers. Our information-theoretic lower bound applies to a general class of causal estimation and manipulation policies for the leader and the follower. We provide a simple learning policy that achieves the lower bound. Finally, we show that under a fixed learning rule, increasing the manipulation budget of the follower can counterintuitively reduce the leader's regret.
\end{abstract}

\section{Introduction}

Incentive design is a hierarchical decision-making problem between a leader and a follower. The leader seeks to induce a desired response from the follower. To that end, the leader commits to an incentive policy that dictates how she would react to the follower's action. Modeled as a Stackelberg game, incentive design finds applications in the design of rule-books such as the tax-code, insurance contracts, etc., e.g., see \cite{ratliff2019perspective, basar1984affine,cansever1982minimum,zheng1982existence,bacsar2024inducement}. While the two players have different strategic incentives, it is commonly assumed that the leader knows \emph{more} than the follower, an assumption that rarely holds in practice. To address that deficit, recent work has allowed the follower to have informational advantage; the follower has a private type that the leader seeks to learn from repeated interactions in order to shape better commitments incentivizing more beneficial behaviors, e.g., in \cite{velicheti2025harnessing,velicheti2026learning}. This paper adds to that growing literature.

Games of asymmetric information have a long history, arising in a variety of contexts, including adverse selection, contract theory, signaling games, and Bayesian persuasion.  \cite{aumann1995repeated, akerlof1970lemons,mussa1978monopoly,milgrom1990rationalizability,spence1978job,kamenica2011bayesian,gentzkow2014costly} define a non-exhaustive list of papers on the topic. Learning through repeated interaction is a natural way to tackle information asymmetry. This approach marries game theory with stochastic approximation, online optimization, and adaptive control literature. Ranging from classical works on the topic such as \cite{hart1985nonzero, aumann1995repeated,sorin2002first,cesa2006prediction,fudenberg1998theory}, the literature on learning in games has specialized to specific contexts, e.g., learning opponent types in Stackelberg setups in \cite{letchford2009learning, balcan2015commitment, amin2013learning, velicheti2024learning}, in Bayesian persuasion or strategic information transmission in \cite{mansour2022bayesian,xu2016signaling, sayin2019hierarchical, castiglioni2020online, babichenko2022regret}, and in multi-agent reinforcement learning in \cite{jin2022power,mao2023provably}, etc. In this paper, we consider a leader-follower quadratic Gaussian incentive design game, where the follower has a private type. The leader must incentivize the follower, but can learn about the follower through repeated interactions. Recently in \cite{velicheti2026learning}, it was shown that an explore-then-commit learning style allows the leader to obtain sub-linear ${\cal O}(\sqrt{T})$ regret in $T$ rounds of interaction when the follower provides a myopic best response to the leader in every round, but it may be in the follower's best interest to  force the leader to attain linear regret if the follower is non-myopic. This work builds on that example game with a non-myopic agent. 

Using the quadratic game studied by \cite{velicheti2026learning}, we allow the follower to manipulate his myopic best response in order to steer the leader's estimates toward values he prefers, thereby sacrificing immediate performance for potential future gains.  The interaction begins with the leader announcing her learning policy, which uses past observations to estimate the follower's private type, and how she will use that estimate in an incentive policy. Knowing how the leader will use these observations, which are noisy realizations of the follower's actions, the follower commits to a manipulation policy from the prescribed class but does not announce it. 
In effect, both the learning and the manipulation policies are represented as causal functions of the observations available to the respective agents in each round. In this setup, we establish minimax regret lower bounds for the leader over this causal policy class under bounded manipulation and approximate follower rationality in which the manipulation budget bounds the expected cumulative squared deviation of the follower’s chosen actions from his myopic best response. This allows us to examine the effect of manipulation, ranging from a fully myopic setting with zero budget to a setting that allows non-myopic behavior under a linearly growing manipulation budget. We then provide a simple learning policy within the same class that achieves the matching minimax regret rate.

If the follower is nonstrategic or fully myopic, this interaction would resemble a classical adaptive-control problem in which the leader seeks to control an environment (with the follower as a responder) some of whose dynamics are initially unknown. Such problems require the tradeoff between exploitation of past data, typically represented by applying a control suited to some type of estimate of the environment, and exploration, which involves gathering information to improve said estimate as given in \cite{guclu2024trail, guclu2026unified, ziemann2021uninformative, ziemann2025regret}. This tradeoff changes fundamentally when the unknown environment is itself a non-myopic strategic agent. The leader cannot assume that her actions will elicit observations that truthfully reveal the follower's private information. \emph{The leader must learn the follower's private type from observations that the follower himself can strategically influence.} In these problems, the follower's manipulation today affects the leader's estimates down the road, which in turn affects the trajectory of the leader's incentive designs, and the resulting action sequence of the follower. Such dynamic interactions of non-myopic agents have been studied by \cite{pavan2014dynamic,amin2013learning,haghtalab2026learning}. While our work shares the same spirit, we differ in that our focus is on the impact of \emph{bounded} manipulation capabilities on the leader's estimation quality and her overall costs. 

Statistical tools used in adaptive control and sequential decision-making have proven useful for studying learning in games. These include parameter-estimation lower bounds based on the van Trees inequality examined in \cite{guclu2024trail, ziemann2021uninformative, ziemann2025regret} and Fano-type information-theoretic methods that relate learning performance to the ability to distinguish between underlying models as given in \cite{lattimore2020bandit, lai1985asymptotically, tsybakov2009introduction}. For example, \cite{ai2024dynamic} study dynamic service-fee pricing with forward-looking buyers and use the multivariate van Trees inequality to establish a regret lower bound. Meanwhile, \cite{ghoshal2017learning, ghoshal2018learning} use Fano's inequality to establish sample-complexity lower bounds for recovering Nash-equilibrium sets in graphical and poly-matrix games from noisy observations of players' actions. We use Fano's inequality and an imitation-based argument between different follower types to establish our lower bounds. The difficulty with applying information-theoretic lower bounds based only on observation noise is that manipulation, while potentially making estimation harder, may also reveal information about the very parameter the follower wishes to conceal, since his manipulation choices depend on his private type. We carefully account for this information conveyed by manipulation itself.

Finally, we ask whether manipulation is inherently harmful to the leader and answer this question negatively. We show an example where the leader's learning algorithm introduces a bias that skews her estimates away from the true parameter. By trying to manipulate the leader, the follower partially offsets this bias, helping the leader estimate better and thus reducing her regret. As the follower's manipulation budget increases further, however, manipulation begins to hurt the leader, producing an interesting non-monotone relationship between the manipulation budget and the leader's performance.

\paragraph*{Paper Organization and Our Contributions:}
We present the interaction model in a static setup in Section \ref{sec:model}, and its dynamic counterpart in Section \ref{sec:dynamicBoundedManip}. Then in Section \ref{sec:leaders_performance}, we (i) establish minimax regret rate for a leader learning a follower's private type, and (ii) we construct a learning policy that matches the regret under approximate rationality of the follower. In Section \ref{sec:manipulation_can_be_helpful} we show that, in some cases, the follower's manipulation can benefit the leader. We conclude the paper with discussions about future research in Section \ref{sec:conclusion}. All proofs are in the appendix.

\section{The Interaction Model}\label{sec:model}
Consider a game between a leader and a follower, where the leader's action is $u \in \mathbb{R}$ and the follower's action is $v\in\mathbb{R}$. 
 The leader's and the follower's costs are, respectively, 
\begin{align}
c_L(u,v;\zeta)
&=(\zeta-u-v)^2+2u^2+\beta v^2,
\label{eq:cp}\\
c_F(u,v;\zeta)
&=(\zeta-u-v)^2+v^2,
\label{eq:ca}
\end{align}
where $\beta>0$ is common knowledge. The costs depend on the parameter $\zeta$. Assume that $
\zeta\in \mathcal{Z}=[-\overline{Z},\overline{Z}],$ with $ \overline{Z}>0$ is the follower's private type that is known to the follower, but unknown to the leader. Define
$
d:=\beta^2+2\beta-2 \neq 0. 
$
In the sequel, we study a $T$-period interaction between the two players, where the leader commits beforehand at every stage, how she would react to the follower's action. Before delving into this multi-stage interaction, we start by examining the one-stage Stackelberg game between them in the case where the leader knows $\zeta$, and uses an affine incentive rule
\begin{equation}
u(v)=av+k\zeta,
\qquad a,k\in\mathbb{R}.
\label{eq:generic-affine}
\end{equation}
The follower observes this rule and chooses $v$ to minimize his own cost by solving
\[
\arg\min_{v\in\mathbb R}
c_F(av+k\zeta,v;\zeta).\]
The leader controls the myopic best-response action of the follower via her announced incentive rule, even though she cannot actually control what action the follower plays. However, if the leader had full control over both actions, she would solve
$
(u^\star(\zeta),v^\star(\zeta))
:=
\arg\min_{(u,v)\in\mathbb R^2}
c_L(u,v;\zeta).$ Since $c_L(\cdot,\cdot;\zeta)$ is strictly convex, this minimizer is unique and is given by
\begin{equation}
u^\star(\zeta)
=
\frac{\beta}{3\beta+2}\zeta,
\qquad
v^\star(\zeta)
=
\frac{2}{3\beta+2}\zeta,
\label{eq:full-info}
\end{equation}
which produces the full-information benchmark cost
\begin{equation}
c_L^\star(\zeta)
:=
c_L\bigl(u^\star(\zeta),v^\star(\zeta);\zeta\bigr)
=
\frac{2\beta}{3\beta+2}\zeta^2.
\label{eq:full-info-cost}
\end{equation}
Without the ability to directly control the follower's action $v$, suppose that the leader commits (before any actions are played) that she would react to the follower by playing via the incentive rule, $u(v)
=
a v+k\zeta$ with 
\begin{equation}
a
=
\frac{1-\beta}{\beta},
\qquad
k
=
\frac{d}{\beta(3\beta+2)}.
\label{eq:implementation-coefficients}
\end{equation}
Under this rule, the follower minimizes his own cost function and produces the myopic response,
$
\frac{2}{3\beta+2}\zeta
=
v^\star(\zeta),
$
and the follower's response yields
$
u\bigl(v^\star(\zeta)\bigr)
=
a v^\star(\zeta)+k\zeta
=
\frac{\beta}{3\beta+2}\zeta
=
u^\star(\zeta).
$
Hence, the incentive mechanism in effect implements the leader's full-information benchmark through the follower's per-stage optimal response. This defines the core of the \emph{indirect mechanism} for solving an equilibrium for incentive design games in \cite{basar1984affine}.

The aforementioned incentive policy is not implementable when the leader does not know the private type $\zeta$ of the follower and can only estimate it from her observations. At each $t$, suppose that the leader observes a noisy version of the follower's action,
\[
Y_t=v_t+\sigma_v\varepsilon_t,
\] for $\sigma_v>0,
$
where $\varepsilon_t \stackrel{\mathrm{i.i.d.}}{\sim} \mathcal N(0,1)$ are standard i.i.d. Gaussian random variables.  Using the history of these observations, suppose that the leader suitably creates an estimate of $\zeta$, denoted by $z_t$, and commits to playing an incentive policy that mimics that in the full-information setting, but replaces $\zeta$ with $z_t$ and $v_t$ by $Y_t$, i.e., 
\begin{equation}
\widetilde u_t
=
a Y_t +k z_t.
\label{eq:leader-static}
\end{equation}
Define $u_t := a v_t + k z_t$, which will be used to simplify the notation in the following analysis. 
If the follower is a myopic best-responder, then he minimizes his cost with respect to this incentive strategy and the resulting response is \footnote{For fixed $z_t$, the follower solves
$
v_t^0
=
\arg\min_{v_t\in\mathbb R} \E_\zeta[
c_F\bigl(a v_t+ a \sigma_v \varepsilon_t + k z_t,v_t;\zeta\bigr)|z_t, v_t]
=
\arg\min_{v_t\in\mathbb R}
\left\{
(\zeta-k z_t-\frac{v_t}{\beta})^2+v_t^2
\right\},
$
where the expectation is taken with respect to $\varepsilon_t$, conditioned on $\zeta$, $v_t$, and $z_t$. The second equality follows from the fact that $\varepsilon_t$ is independent of \(z_t\) and the follower’s action and has zero mean and unit variance. 
}
$
A\zeta+B z_t,
$
where
\begin{equation}
A:=\frac{\beta}{\beta^2+1},
\qquad
B:=-\frac{d}{(3\beta+2)(\beta^2+1)}.
\label{eq:AB}
\end{equation}

Our goal is to study how the follower can systematically manipulate the leader's observations over time, shaping her estimate, which might benefit the follower down the line. To study such strategic interaction, we allow the follower to play
\begin{equation}
v_t=A\zeta+B z_t+m_t,
\label{eq:v-decomp}
\end{equation}
where $m_t$ is the follower's deviation from or the manipulation injected into his  myopic best response. We begin our analysis by studying if the follower cares about such deviation.

\emph{Why should the follower manipulate?}
Substituting
\eqref{eq:v-decomp}
and the leader's incentive rule into the follower's stage cost gives
\begin{equation}
\begin{aligned}
\mathbb E_\zeta
[c_F(\widetilde u_t,v_t;\zeta)\mid z_t,v_t]
&=
c_F(u_t,v_t;\zeta)+a^2\sigma_v^2 =
D m_t^2
+q\bigl(z_t-z^\star(\zeta)\bigr)^2
+a^2\sigma_v^2,
\end{aligned}
\label{eq:follower-square}
\end{equation}
where the expectation is taken with respect to the observation noise $\varepsilon_t$ conditioned on $\zeta$ and 
\begin{equation}
q
:=
\frac{d^2}
{(3\beta+2)^2(\beta^2+1)},
\qquad
z^\star(\zeta)
:=
\frac{\beta(3\beta+2)}{d}\zeta.
\label{eq:agent-reference}
\end{equation}
The derivation can be found in the Appendix. Notice that $\zeta \neq z^\star(\zeta)$ for any $\zeta \neq 0$. Since $z^\star(\zeta)=\zeta$ would require $\beta(3\beta+2)=\beta^2+2\beta-2$, equivalently $\beta^2=-1$, which is impossible. This simple calculation showcases the central trade-off for the follower. Deviating from his myopic action via the addition of $m_t$ incurs a cost $\simeq m_t^2$, while the second term of the right-hand side of \eqref{eq:follower-square} is minimized by driving the leader's type estimate towards $z^\star(\zeta)$.

Recall that if the type estimate equals the type itself, i.e., $z_t=\zeta$, then 
$
u_t(v_t)=a v_t + k z_t
$
coincides with the optimal full-information incentive mechanism, and  the leader commits to the policy, $\widetilde u_t = u_t(v_t) + a \sigma_v \varepsilon_t$. Notice that 
\begin{align}
\begin{aligned}
\E_\zeta\left[c_L(\widetilde u_t,v_t;\zeta)|v_t, z_t\right]
&=\E_\zeta\left[(\zeta-\widetilde u_t-v_t)^2+2\widetilde u_t^2|v_t, z_t \right]+\beta v_t^2 \\
&= (\zeta- u_t-v_t)^2+2 u_t^2+\beta v_t^2 + 3a^2 \sigma_v^2\\
& = c_L(u_t,v_t;\zeta) + 3a^2 \sigma_v^2.
\end{aligned}\label{cost_of_leader}
\end{align}
The second term of the last equality is not policy-dependent. Hence, in the absence of manipulation, the leader's expected stage cost is minimized at \(z_t=\zeta\), whereas the follower’s tracking term is minimized at \(z_t=z^\star(\zeta)\). 

\section{Dynamic Interaction with Bounded Manipulation}
\label{sec:dynamicBoundedManip}
Now, we consider the case where the leader and the follower interact over $T$ periods, where the observations of actions from the follower shape the leader's estimate of the former's type, while the follower deviates (in a bounded fashion as delineated below) from his myopic best response to manipulate the leader's estimation process.
Before the game starts, let the leader announce her policy, which consists of two parts: the incentive structure in \eqref{eq:leader-static}, whose constants remain fixed throughout the horizon, and the learning policy, which produces the estimate $z_t$ from past observations $(Y_1,\dots,Y_{t-1})$ as well as the history of the estimates $(z_1,\dots, z_{t-1})$. In other words, the leader commits to an incentive policy structure and how she will use the observations to form her estimate of the follower's private type. {We deliberately fix the incentive rule to focus on the leader's learning architecture and how it handles manipulation.}

After observing the leader's commitments, we allow  the follower to choose a manipulation policy that causally determines $m_t$ from the history of estimates $(z_1,\dots,z_t)$, his private type $\zeta$, and his own past actions $(v_1,\dots,v_{t-1})$. The follower does not announce this function. At each stage, the leader produces $z_t$ using the announced learning policy and allows the follower to know $z_t$. The follower then chooses $v_t$ through his private but committed manipulation policy, together with \eqref{eq:v-decomp}. Finally, the leader observes the noisy version, $Y_t$ and implements $\widetilde u_t$ according to the announced incentive rule in \eqref{eq:leader-static}. 

We start by defining the policy class of the leader. At time \(t\), she uses \((Y_s)_{s<t}\) and \((z_s)_{s<t}\), to produce the estimate \(z_t\). We assume $(\varepsilon_t)_{t = 1:T}$ are independent of $\zeta$. The leader implements the incentive rule
given in \eqref{eq:leader-static}, hence, the incentive structure is not considered part of the leader's policy class given as
\begin{equation}
\label{eq:principal-class}
\mathcal{S}^{\pi}
:=
\left\{
\pi=((\pi_t)_{t=1}^T) \ \mid \
\pi_t : (Y_1,\ldots,Y_{t-1},z_1, \dots, z_{t-1})\mapsto z_t \in \mathcal{Z},\; \forall 1 \leq t \leq T.
\right\}.
\end{equation}
The maps \((\pi_t)_{t=1}^T\) are announced before the game starts, so the follower knows the learning policy before choosing his manipulation function.

Next, we examine the follower's policy set in which the manipulation trajectory is bounded via an \emph{expected} budget, 
\begin{equation}
\begin{aligned}
\mathcal{S}^{\mu}(C_T;\zeta, \pi)
&:=\left\{
\mu=(\mu_t)_{t=1}^T \ \mid \
 \mu_t : (\zeta,(z_s)_{s\leq t},(v_s)_{s< t}) \mapsto m_t, 
 D\mathbb E_\zeta^{\pi, \mu}\sum_{t=1}^Tm_t^2\le C_T
\right\},
\end{aligned}
\label{eq:agent-set}
\end{equation}
with $D:=\frac{\beta^2+1}{\beta^2}$ and $E_\zeta^{\pi, \mu}$ denoting the expectation conditioned on the private parameter being $\zeta$ and the manipulation policy being picked as $\mu$  with the leader following policy \(\pi\). This set is nonempty as zero manipulation is feasible. A strategic follower picks $\mu$ seeking to minimize the cumulative cost given as
\begin{align}
\begin{aligned}
J_F(\mu;\zeta,\pi)&:=
\mathbb E_\zeta^{\pi,\mu}
\!\left[
\sum_{t=1}^T c_F(\widetilde u_t,v_t;\zeta)
\right] =
D\mathbb E_\zeta^{\pi,\mu}\|m\|_2^2
+
q
\mathbb E_\zeta^{\pi,\mu}
\!
\left\|
z^{\pi,\zeta,\mu}
-
z^\star(\zeta)\mathbf 1_T
\right\|_2^2
 + a^2 \sigma_v^2 T,
\label{follower_entire_cost}
\end{aligned}
\end{align}
where $m = [m_1, \dots, m_T]^\top$ denotes the realized manipulation trajectory generated by the feedback policy $\mu$, $\pi$ is the leader's policy, and $z^{\pi,\zeta,\mu}$ is the realized trajectory for $z_t$. For any
$\eta_T>0$, define the class of approximately rational causal feedback
responses by
\begin{equation}
\begin{aligned}
\operatorname{BR}^{\eta_T}(\zeta,\pi,C_T)
:=\bigg\{\mu\in\mathcal{S}^{\mu}(C_T;\zeta, \pi):
J_F(\mu;\zeta,\pi)
\le\inf_{\psi\in\mathcal{S}^{\mu}(C_T;\zeta, \pi)}
J_F(\psi;\zeta,\pi)+\eta_T\bigg\}.
\end{aligned}
\label{eq:approx-best-response}
\end{equation}
Since zero manipulation is feasible and yields a finite
cumulative follower cost by the boundedness of $z_t$
in \eqref{eq:principal-class}, and the objective is
nonnegative, the infimum is finite. Consequently, the
approximate best-response set is nonempty for every
$\eta_T>0$. Every
$\mu\in\operatorname{BR}^{\eta_T}(\zeta,\pi,C_T)$
satisfies, for each feasible alternative $\widehat\mu$,
\begin{equation}
J_F(\mu;\zeta,\pi)
\le J_F(\widehat\mu;\zeta,\pi)+\eta_T.
\label{eq:approx-optimality}
\end{equation}

The leader's \emph{regret} is defined as the cumulative opportunity cost compared to the full-information benchmark 
$c_L^\star(\zeta) + 3 a^2 \sigma_v^2 =2\beta\zeta^2/(3\beta+2) + 3 a^2 \sigma_v^2$ developed in \eqref{cost_of_leader},
\begin{equation}
\begin{aligned}
R_T^L(\zeta,\pi,\mu)
&:=
\mathbb E_\zeta^{\pi,\mu}
\left[
\sum_{t=1}^T
\left(
c_L(\widetilde u_t,v_t;\zeta)
-c_L^\star(\zeta)
-3a^2\sigma_v^2
\right)
\right]= 
\mathbb E_\zeta^{\pi,\mu}
\left[
\sum_{t=1}^T
\left(
c_L(u_t,v_t;\zeta)-c_L^\star(\zeta)
\right)
\right].
\end{aligned}
\end{equation}
Define $e_t = z_t - \zeta$ and $e := [e_1, \dots, e_T]^\top \in \mathbb R^T$. Our results in Section~\ref{sec:leaders_performance} rely on the following result that shows that the leader's regret can be resolved into estimation error and manipulation energy.

\begin{lemma}
\label{lem:regret-output}
For constants $0<\lambda\le\Lambda<\infty$ independent of $T$, 
\begin{equation}
\lambda(\E_\zeta^{\pi,\mu}\|e\|_2^2+\E_\zeta^{\pi,\mu}\|m\|_2^2)
\le R_T^L
\le\Lambda(\E_\zeta^{\pi,\mu}\|e\|_2^2+\E_\zeta^{\pi,\mu}\|m\|_2^2).
\label{eq:regret-equivalence}
\end{equation}
\end{lemma}

\section{Characterization of Leader Performance}\label{sec:leaders_performance}
In this section, we establish a minimax lower bound on the performance of the leader under bounded manipulation and approximate rationality from the follower in Theorem \ref{thm:rational-lower-bound} and then construct a leader's policy that achieves that lower bound in Theorem \ref{cor:minimax-approx}.
We use the notation
$[p]_{\underline p}^{\overline p}
:=\max\{\underline p,\min\{p,\overline p\}\}$.

\begin{theorem}
\label{thm:rational-lower-bound}$
\sup_{\zeta\in\mathcal{Z}} \;  \inf _{\mu\in\operatorname{BR}^{\eta_T}
(\zeta,\pi,C_T)}R_T^L(\zeta,\pi,\mu)
\gtrsim [C_T]_{\log T}^{T}$ for any
$
\eta_T\lesssim \sqrt{T [C_T]_0^T},
$
$\pi\in\mathcal{S}^{\pi}$.
\end{theorem}

In particular, when \(C_T=\Theta(T)\), the theorem permits an approximate rationality tolerance \(\eta_T=\Theta(T)\). This means that the lower bound continues to hold even when the follower is allowed a non-vanishing rationality gap, \(\eta_T/T=\Theta(1)\). Thus, the result does not require the follower to act increasingly more rationally as the horizon grows.

The key idea of the proof is that a leader with a learning policy allowing her to differentiate between two nearby types accurately can create an opportunity for profitable imitation by the follower. We choose types $\zeta_0$ and $\zeta_1$ so that $\zeta_1$ is closer to the type-$\zeta_0$ follower's preferred estimate $z^\star(\zeta_0)$. Suppose that the leader achieves sufficiently small regret for both types under approximately optimal follower responses. By Lemma~\ref{lem:regret-output}, both the expected cumulative estimation error of the leader and the expected manipulation energy of the follower must then be small for each type. The type-$\zeta_0$ follower can imitate the type-$\zeta_1$ response by adding a constant offset to that type's manipulation rule. This generates identical observations and hence identical leader estimates under the same noise realizations. Because the leader learns $\zeta_1$ accurately, imitation moves her estimate toward $\zeta_1$, which the type-$\zeta_0$ follower prefers. We choose the type separation so that this imitation satisfies the manipulation budget and reduces the follower's cumulative expected cost by more than $\eta_T$ relative to his original response. This contradicts the approximate optimality of that response. Consequently, the leader cannot achieve such small regret for both types. 

The preceding argument gives an $\Omega(\min\{C_T,T\})$ lower bound on the leader's regret. The logarithmic lower bound reflects the statistical difficulty of learning an unknown Gaussian mean. Without manipulation, the classical minimax mean-squared estimation error after $t$ observations of i.i.d. Gaussian distributions is $ \simeq 1/t$, giving a logarithmic cumulative estimation-error. In our setting, however, manipulation makes the argument more delicate because it can itself convey information about the unknown parameter. We account for this additional information by bounding its effects via a function of regret itself and show that the additional information conveyed by manipulation cannot reduce the worst-case regret below $\log T$ via Fano's inequality and Lemma \ref{lem:regret-output}.

Next, we construct a leader's policy for which the regret matches the minimax lower bound, thereby becoming an order-optimal policy. 
Consider $\pi^{\mathrm A}$, defined by
\begin{equation}
z_t 
=
\left[\frac{1}{A t}
\sum_{s=1}^{t-1}(
Y_s-B z_s)\right]_{-\overline{Z}}^{\overline{Z}},
\qquad
t=1,\ldots,T.
\label{eq:continual-updating}
\end{equation}
Projection ensures that $z_t\in\mathcal Z$ and by construction, $\pi^A\in\mathcal{S}^{\pi}$. We characterize the regret performance of this estimation policy in the next result.

\begin{theorem}
\label{cor:minimax-approx}
$\sup_{\zeta\in\mathcal{Z}}
\sup_{\mu\in\operatorname{BR}^{\eta_T}
(\zeta,\pi^{\mathrm A},C_T)}
R_T^L(\zeta,\pi^{\mathrm A},\mu)
\lesssim [C_T]_{\log T}^{T}$ for any $C_T\ge0$, $\eta_T\lesssim T$.
\end{theorem}

The upper bound permits an approximate-rationality tolerance $\eta_T=O(T)$, whereas Theorem~\ref{thm:rational-lower-bound} requires $\eta_T\lesssim\sqrt{T\min\{C_T,T\}}$. These conditions arise for different reasons. For the upper bound, approximate rationality is used to control the follower's cumulative squared manipulation energy by $O(T+\eta_T)$, so $\eta_T \lesssim T$ suffices for the result. The lower-bound argument instead relies on imitation between two types improving the follower's cost by more than $\eta_T$ relative to an $\eta_T$-optimal policy, contradicting the approximate optimality. Therefore, the allowable values of $\eta_T$ are more restricted.

Our results characterize the impact of the manipulation budget \(C_T\) on the leader's performance. In this sense, they bridge learning with fully myopic followers and the linear-regret phenomena associated with unrestricted non-myopic behavior, as discussed in \cite{haghtalab2026learning, velicheti2026learning}. We remark that learning from observations shaped by strategic agents also arises in strategic classification, where individuals modify their observable features, while paying a manipulation cost, to obtain more favorable classification outcomes as examined in \cite{hardt2016strategic,dong2018strategic}. While the main theoretical guarantees of \cite{hardt2016strategic} assume access to the agents' manipulation cost functions, \cite{dong2018strategic} show that learning is possible without observing either the individuals' original features or their manipulation cost functions.  In the same spirit, Theorem~\ref{cor:minimax-approx} shows that \emph{the leader does not need to know how much manipulation the follower is capable of to achieve order-optimal regret.}

\section{Can Manipulation Benefit the Leader?}\label{sec:manipulation_can_be_helpful}
One might surmise that increasing the manipulation budget of the follower would hurt the leader's performance. We now demonstrate that it can be exactly the opposite! We consider the case where the follower can have one of two private types $\zeta = -1$ and $\zeta = 1$, whereas the leader commits to the learning rule $ \pi: z_t = Y_{t-1}-B z_{t-1}$ with $ z_1 = 0.$ For this example, we relax the restriction $z_t\in\mathcal Z$ for learning algorithms and allow the leader's estimates to take values in $\mathbb R$. Under the learning rule $\pi$, the estimates satisfy $ z_{t+1} = A\zeta+m_t+\sigma_v\varepsilon_t. $ In the absence of manipulation, we have $ \mathbb E_{\zeta}^{\pi}[z_t]=A\zeta $ for $t\ge2$, with the expectation taken with respect to the observation noise, which need not equal the true type $\zeta$. When $0\leq A<1$, the leader's estimate is biased toward zero. On the other hand, the follower prefers the leader's estimate to approach $z^\star(\zeta)$. When $z^\star(\zeta)$ lies beyond $\zeta$ in the direction away from $A\zeta$, the follower's manipulation can move $z_t$ in a direction that offsets the estimation bias $\zeta(A-1)$ induced by the learning rule. This occurs for $\beta=3/2$, which gives $A=6/13$ and $z^\star(\zeta)=3\zeta$. Consequently, a small amount of manipulation can move the estimate closer to the true type, decreasing the estimation error, even though the follower's objective is to steer it toward $z^\star(\zeta)$. However, as the manipulation budget increases, the direct cost of manipulation to the leader can eventually outweigh the improvement in estimation accuracy, as captured by the regret lower bound established in Lemma~\ref{lem:regret-output}, where one term captures the expected cumulative estimation error and the other term captures the effect of manipulation on the leader's cost. Thus, increasing the manipulation budget can initially decrease the leader's regret but eventually increase it, resulting in a nonmonotone relationship between the manipulation budget and the leader's performance.

 Before the game starts, the follower knows leader's learning rule, and selects a causal manipulation policy satisfying the  expected-budget constraint  in \eqref{eq:agent-set} with $C_T = cT$ for some $c \geq 0$. We will show that the optimal response $\mu^\star \in \arg\min_{\psi\in\mathcal{S}^{\mu}(c T;\zeta,\pi)}
J_F(\psi;\zeta,\pi)$ exists and study
\begin{equation}
V_{\beta}(c):= \lim_{T \to \infty}\max_{\zeta\in\{-1,1\}}
\frac1T R_T^L(\zeta,\pi,\mu^\star).
\label{eq:tri-value}
\end{equation}
A decrease in $V_{\beta}(c)$ represents an improvement in the leader's worst-case performance.

We begin by characterizing the follower's optimal manipulation rule $\mu^\star$ under
the learning rule $\pi$. Recall from \eqref{eq:follower-square} that for any manipulation rule $\mu$, 
the follower's expected stage cost is
$
\mathbb E_\zeta^{\pi,\mu}
[c_F(\widetilde u_t,v_t;\zeta)]
=
D\mathbb E_\zeta^{\pi,\mu}[m_t^2]
+
q\mathbb E_\zeta^{\pi,\mu}
[(z_t-z^\star(\zeta))^2]
+
a^2\sigma_v^2, 
$
where $z^\star(\zeta)=\frac{\beta(3\beta+2)}{d}\zeta.
$ Since $z_1=0$, and $m_t$ is independent of the current
observation noise $\varepsilon_t$, the follower's cumulative
expected cost becomes
\begin{align}
J_F(\mu;\zeta,\pi)
=&
q
\left(
\frac{\beta(3\beta+2)}{d}
\right)^2\zeta^2
+
(T-1)q\sigma_v^2  + T a^2 \sigma_v^2
\nonumber\\
&+
\sum_{t=1}^{T-1}
\mathbb E_{\zeta}^{\pi,\mu }\left[
D m_t^2
+
q
\left(
(A\zeta+m_t)
-
\frac{\beta(3\beta+2)}{d}\zeta
\right)^2
\right]
+
D\mathbb E_{\zeta}^{\pi,\mu }[m_T^2].
\label{eq:tri-follower-cost}
\end{align}
Notice that the first three terms do not depend on the
follower's actions and manipulation strategy. Consequently, $\mu^\star$ can be obtained by minimizing the
remaining terms subject to the budget constraint.  In the final period, manipulation incurs a
quadratic cost without affecting any future estimate,
so $m_T^\star=0$. Calculating the manipulation strategy for the remaining time steps boils down to minimizing the fourth term. For a convex function $f$, and any feasible policy $\mu$, Jensen's inequality gives
$
\sum_{t=1}^{T-1}
\mathbb E_\zeta^{\pi,\mu}[f(m_t)]
\ge (T-1)f(\bar m) $ and $
\bar m:=\frac1{T-1}
\sum_{t=1}^{T-1}
\mathbb E_\zeta^{\pi,\mu}[m_t].$
\begin{figure}[t]
    \centering
    \includegraphics[width=0.5\linewidth]{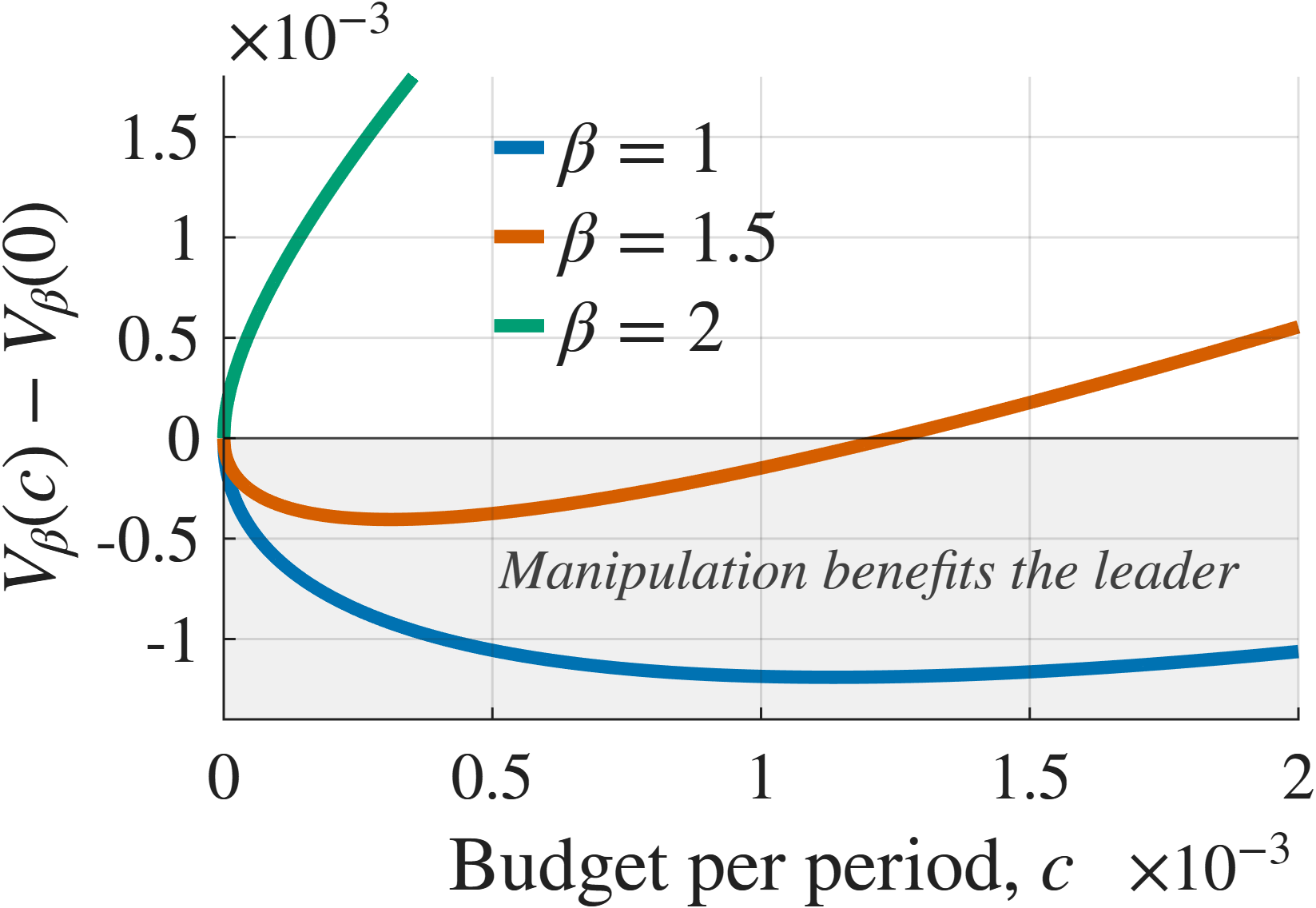}
    \caption{Effect of the manipulation budget on the leader's
    limiting average regret for different values of $\beta$,
    with  $\sigma_v=0.3$. 
    $V_\beta(c)-V_\beta(0) < 0 \implies$ manipulation benefits
    the leader, which is visible for $\beta \in \{1, 1.5\}$.}
    \label{fig:manipulation-benefit}
\end{figure}
Thus, although the manipulation $m_t$ may vary across periods, each summand of the fourth term of $J_F(\mu;\zeta,\pi)$ is the same strictly convex quadratic function of $m_t$ at every $t=1,\ldots,T-1$. Therefore, replacing any causal manipulation policy with the deterministic choice $m_t=\bar m$ for all $t=1,\ldots,T-1$ cannot increase the follower's cost. Moreover, Jensen's inequality also gives
$
(T-1)\bar m^2
\le
\sum_{t=1}^{T-1}\mathbb E_{\zeta}^{\pi,\mu}[m_t^2],
$
so this replacement does not violate the manipulation
budget if the original trajectory already follows this budget.  Consequently, the optimal manipulation
is deterministic and constant over the first $T-1$
periods. Using this, the follower's exact manipulation strategy can be formulated as
$
m_t^\star
=
\zeta\,\kappa_{T}(c),$ for all $ t=1,\ldots,T-1$ and $m_T^\star = 0$
where
\begin{equation}
\kappa_T(c)
=
\left[
\frac{q(k^{-1}-A)}{D+q}
\right]_{-\sqrt{\frac{cT}{D(T-1)}}}^{\sqrt{\frac{cT}{D(T-1)}}}.
\label{eq:tri-mu}
\end{equation}
Fix $\beta=3/2$ and $\sigma_v=3/10$.
For $0\le c\le0.01$ and $T\ge2$, the manipulation
budget is binding,  which results in
$\kappa_T(c)=\sqrt{9cT/[13(T-1)]}$.
Substituting the follower's optimal response into the
leader's regret expression and taking $T\to\infty$ gives
\begin{equation}
V_{3/2}(c)
=
\frac{6421}{43940}
-
\frac{28}{507}\sqrt{\frac{9c}{13}}
+
\frac{441}{338}c,
\qquad 0\le c\le0.01.
\label{eq:tri-explicit-V}
\end{equation}
In particular, $V_{3/2}(0.0003)<V_{3/2}(0)<V_{3/2}(0.01)$,
showing that increasing the manipulation budget
can initially benefit the leader but subsequently
increase its regret. As explained above, this comes from the estimation bias of the leader's learning rule partially being offset by the manipulation of the follower. Without manipulation, $\mathbb E_\zeta^\pi[z_t]=A\zeta\approx0.462\zeta$ for $t\ge2$. With $c=0.0003$ and sufficiently large $T$, the follower's optimal manipulation gives $\mathbb E_\zeta^{\pi,\mu^\star}[z_t] =(A+\kappa_T(c))\zeta\approx0.476\zeta$. This moves the estimate closer to the true type, reducing both estimation error and regret.

In Figure~\ref{fig:manipulation-benefit}, we fix $\sigma_v=0.3$ and compare $\beta\in\{1,1.5,2\}$ under the same learning rule $\pi$. We restrict attention to $0\le c\le0.002$. The horizontal axis shows the manipulation budget per period, $c$, and the vertical axis shows $V_\beta(c)-V_\beta(0)$, the change in the leader's limiting average regret relative to zero manipulation at the same value of $\beta$. Consequently,  $V_\beta(c)-V_\beta(0) < 0$ indicates that manipulation benefits the leader, which is visible for $\beta\in\{1,1.5\}$ where the regret initially decreases due to the follower's manipulation partially correcting the bias in the leader's estimates. However, after a point, increasing the manipulation budget increases regret. On the other hand, for $\beta=2$, manipulation increases the leader's regret from the start. Thus, whether manipulation benefits the leader depends on both $c$ and $\beta$.

\section{Conclusion} 
\label{sec:conclusion}
In this paper, we study a repeated quadratic Gaussian Stackelberg game, where the follower knows a private parameter governing both players' costs, as well as the leader's learning rule, which determines how the leader uses the past observations to estimate said follower's private parameter. We establish regret lower bounds over a general class of causal learning policies using a budget accounting for the follower's ability to strategically manipulate the observations. We further show that a simple learning policy achieves the matching minimax order under approximate follower rationality. Finally, we show that manipulation is not inherently harmful to the leader. We give an example where the follower's manipulation partially corrects an estimation bias present in the leader's learning architecture and thereby reduces the leader's regret. 

Although our analysis focuses on a specific class of quadratic Stackelberg games, we believe that the underlying ideas may extend to more general strategic interactions with information asymmetry. In future work, we aim to study more general settings where agents may have different private information, and their learning and action-selection policies are causal mappings of past observations and may be randomized. In these problems, our goal is to establish both minimax regret lower bounds and to characterize learning rules that achieve these bounds.

\section*{Acknowledgments}
This work was partially supported by the U.S. National Science Foundation under Grant No. ECCS-EPCN 2349418. 

We used ChatGPT and Claude to assist with mathematical calculations, coding, and proofreading the manuscript. The authors assume full responsibility for the accuracy of the final manuscript.

\bibliographystyle{plainnat}
\bibliography{references_checked}

\appendix

\section{Derivation of \eqref{eq:follower-square}}
Let
$r_t:=\zeta-k z_t$. Since
$A\zeta+B z_t=\beta r_t/(\beta^2+1)$,
\eqref{eq:v-decomp} implies
$
v_t=\frac{\beta}{\beta^2+1}r_t+m_t$ and $
\zeta-u_t-v_t
=
\frac{\beta^2}{\beta^2+1}r_t-\frac{m_t}{\beta}.
$
Hence, we have
\begin{align}
c_F(u_t,v_t;\zeta)=
\left(
\frac{\beta^2}{\beta^2+1}r_t-\frac{m_t}{\beta}
\right)^2
+
\left(
\frac{\beta}{\beta^2+1}r_t+m_t
\right)^2 =
\frac{\beta^2}{\beta^2+1}r_t^2
+
\frac{\beta^2+1}{\beta^2}m_t^2.
\label{eq:app-agent-square-one}
\end{align}
Since
$r_t=-k\bigl(z_t-z^\star(\zeta)\bigr)$ and
$\beta^2k^2/(\beta^2+1)=q$,  \eqref{eq:app-agent-square-one}
gives
\[
c_F(u_t,v_t;\zeta) =
D m_t^2
+q\bigl(z_t-z^\star(\zeta)\bigr)^2.
\]
Since $\widetilde u_t=u_t+a\sigma_v\varepsilon_t$
and $\varepsilon_t$ is independent of $(z_t,v_t)$,
with zero mean and unit variance, we have
$
\mathbb E_\zeta
[c_F(\widetilde u_t,v_t;\zeta)\mid z_t,v_t]
=
c_F(u_t,v_t;\zeta)+a^2\sigma_v^2,
$
which yields \eqref{eq:follower-square}.

\section{Proof of Lemma \ref{lem:regret-output}}
Recall that $v_t=A\zeta+Bz_t+m_t$ and $z_t=\zeta+e_t$.
Since $A+B=2/(3\beta+2)$, we have
\[
v_t
=(A+B)\zeta+Be_t+m_t
=v^\star(\zeta)+Be_t+m_t.
\]
Moreover, the full-information benchmark actions satisfy
$
a v^\star(\zeta)+k\zeta
=
u^\star(\zeta).
$
Therefore,
\[
u_t
=av_t+kz_t =u^\star(\zeta) + kz_t - k \zeta + a B e_t + a m_t = u^\star(\zeta)+(aB+k)e_t+am_t.
\]
Thus, the deviations from the full-information benchmark
actions are
\begin{align}
\Delta v_t
&:=v_t-v^\star(\zeta)=Be_t+m_t,\nonumber\\
\Delta u_t
&:=u_t-u^\star(\zeta)=(aB+k)e_t+am_t.
\label{delta_uv_linear}
\end{align}
The leader's stage cost is quadratic and its gradient
vanishes at $(u^\star(\zeta),v^\star(\zeta))$.
Thus, its second-order Taylor expansion is exact which gives
\begin{align}
c_L(u_t,v_t;\zeta)-c_L^\star(\zeta)
=
3(\Delta u_t)^2
+2\Delta u_t\Delta v_t
+(1+\beta)(\Delta v_t)^2.
\label{eq:cost_taylor}
\end{align}
Both action deviations are linear in the estimation
error and manipulation. Substituting
\eqref{delta_uv_linear} to \eqref{eq:cost_taylor} therefore gives
\begin{equation}
c_L(u_t,v_t;\zeta)-c_L^\star(\zeta)
=
\begin{bmatrix}e_t\\m_t\end{bmatrix}^{\!\top}
Q_\beta
\begin{bmatrix}e_t\\m_t\end{bmatrix},
\label{eq:app-Q-form}
\end{equation}
where
\[
Q_\beta:=
\begin{bmatrix}
\displaystyle
\frac{(\beta+1)d^2}
{(3\beta+2)(\beta^2+1)^2}
&
\displaystyle
-\frac{(\beta-1)d}{\beta(\beta^2+1)}
\\[3mm]
\displaystyle
-\frac{(\beta-1)d}{\beta(\beta^2+1)}
&
\displaystyle
\frac{\beta^3+2\beta^2-4\beta+3}{\beta^2}
\end{bmatrix}.
\]
Since $\beta>0$ and $d\neq0$, its first diagonal
entry is positive and
\[
\det Q_\beta
=
\frac{d^2}{\beta^2(3\beta+2)}>0.
\]
Hence, $Q_\beta$ is positive definite. Define $\lambda:=\lambda_{\min}(Q_\beta)>0$
and $\Lambda:=\lambda_{\max}(Q_\beta)<\infty$. Then, \eqref{eq:app-Q-form} implies
\begin{align}
\lambda(e_t^2+m_t^2)
\le
c_L(u_t,v_t;\zeta)-c_L^\star(\zeta)
\le
\Lambda(e_t^2+m_t^2).
\label{eq:one_step_bound}
\end{align}
The implemented incentive $\widetilde u_t$ adds
$3a^2\sigma_v^2$ to the expected stage cost such that 
\begin{align}
\E_\zeta^{\pi, \mu}\left[c_L(\widetilde u_t,v_t;\zeta)|u_t, v_t\right]
- c_L(u_t,v_t;\zeta) = 3a^2 \sigma_v^2.
\label{noise_offset}
\end{align}
Consequently, summing \eqref{eq:one_step_bound} over $t$, taking expectations, using \eqref{noise_offset}, and canceling the noise offset proves the result.

\section{Proof of Theorem \ref{thm:rational-lower-bound}}

The proof has two parts. We first establish an $\Omega([ C_T]_0^T)$ lower bound, and then establish an $\Omega(\log T)$ lower bound for the same quantity.

\subsection{Proving the $\Omega([ C_T]_0^T)$ lower-bound}
Set $\zeta_0 := \overline{Z}/2 \in \mathcal{Z}$ and
\[
\zeta_1 := \zeta_0 + c_0 \operatorname{sgn}(z^\star(\zeta_0) - \zeta_0) \sqrt{\min \{ C_T, T\} / T}
\]
\[
\text{ with } c_0
:=
\min\left\{
\frac{\overline{Z}}{4},\,
\frac{|\zeta_0-z^\star(\zeta_0)|}{4},\,\frac{1}{
\sqrt{D}\left(|A|+\frac1{16}\right)},\,
\frac{
q\left|\zeta_0-z^\star(\zeta_0)\right|
}{
4D\left(|A|+\frac{1}{16}\right)^2
}
\right\}.
\]
Here, the sign function is well-defined since $\zeta_0 \neq z^\star(\zeta_0)$. The first condition ensures that $\zeta_1 \in \mathcal{Z}$, whereas the second ensures that $\zeta_1$ lies between $\zeta_0$ and $z^\star(\zeta_0)$. Fix a leader policy $\pi\in\mathcal{S}^{\pi}$ and choose arbitrary
$\mu_\star^i\in\operatorname{BR}^{\eta_T}(\zeta_i,\pi,C_T)$
for $i\in\{0,1\}$. Define
$z^i:=(z_1^{\pi,\zeta_i,\mu_\star^i},\ldots,z_T^{\pi,\zeta_i,\mu_\star^i})^\top$
as the estimate trajectory generated under leader policy
$\pi$ and follower policy $\mu_\star^i$ at type $\zeta_i$.
Let $m_\star^i:=(m_{\star,1}^i,\ldots,m_{\star,T}^i)^\top$
and $e^i:=z^i-\zeta_i\mathbf1_T$ denote the corresponding
manipulation and estimation-error trajectories, respectively.

Suppose, toward a contradiction, that
\begin{equation}
R_T^L(\zeta_i,\pi,\mu_{\star}^{i})
<
\frac{\lambda c_0^2}{256}
\min\{C_T,T\},
\qquad i=0,1,
\label{eq:small-regret-revised}
\end{equation}
where $\lambda$ is the constant given in
Lemma~\ref{lem:regret-output}. We will construct a causal manipulation policy $\widehat \mu$ for the $\zeta_0$- agent such that $\widehat\mu\in
\mathcal{S}^{\mu}(C_T;\zeta_0,\pi)$ and $J_F(\mu_{\star}^0;\zeta_0,\pi)
-J_F(\widehat\mu;\zeta_0,\pi) \ge 2\eta_T>\eta_T$. However, since $\eta_T > 0$, this contradicts the $\eta_T$-rationality of the policy $\mu_\star^0$ for the $\zeta_0$-agent.

To that end, notice that Lemma \ref{lem:regret-output} and \eqref{eq:small-regret-revised} implies that
\begin{equation}
\begin{aligned}
\mathbb{E}_{\zeta_i}^{\pi,\mu_{\star}^{i}}
\left[\|e^i\|_2^2\right]
\le
\frac{c_0^2}{256}\min\{C_T,T\},\quad
\mathbb{E}_{\zeta_i}^{\pi,\mu_{\star}^{i}}
\left[\|m_{\star}^{i}\|_2^2\right]
\le
\frac{c_0^2}{256}\min\{C_T,T\}.
\end{aligned}
\label{eq:small-errors-revised}
\end{equation}
Define $\widetilde m_t^1:=\mu_{\star,t}^1(\zeta_1,z_{1:t},v_{1:t-1})$
and $\widetilde m^1:=(\widetilde m_1^1,\ldots,\widetilde m_T^1)^\top$.
The causal policy $\widehat\mu$ for the type-$\zeta_0$ follower is picked as
$\widehat m_t:=\widetilde m_t^1+A(\zeta_1-\zeta_0)$. Let $\widehat m:=(\widehat m_1,\ldots,\widehat m_T)^\top$
denote the resulting manipulation trajectory. Since $A\zeta_0+\widehat m_t=A\zeta_1+\widetilde m_t^1$,
the two followers produce the same action under the same history.
Fixing the same observation noises, their initial estimates
coincide, and induction gives identical action, observation, and estimate
trajectories because the leader uses the same learning maps. Consequently,
\begin{equation}
z^{\pi,\zeta_0,\widehat\mu}
=
z^{\pi,\zeta_1,\mu_\star^1}
=
\zeta_1\mathbf{1}_T+e^1
\text{ pathwise under the same
$\{\varepsilon_t\}_{t=1}^T$
}.
\label{eq:imitation-revised}
\end{equation}

Next, we show that
$\widehat\mu\in
\mathcal{S}^{\mu}(C_T;\zeta_0,\pi)$. To that end,
use \eqref{eq:small-errors-revised} and
Minkowski's inequality to obtain
\begin{align}
\left(
\mathbb{E}_{\zeta_0}^{\pi,\widehat\mu}
\|\widehat m\|_2^2
\right)^{1/2}
&=
\left(
\mathbb{E}_{\zeta_0}^{\pi,\widehat\mu}
\left\|
\widetilde m^1+
A(\zeta_1-\zeta_0)\mathbf{1}_T
\right\|_2^2
\right)^{1/2}
\nonumber\\
&\le
\left(
\mathbb{E}_{\zeta_0}^{\pi,\widehat\mu}
\|\widetilde m^1\|_2^2
\right)^{1/2}
+
|A|\,|\zeta_1-\zeta_0|
\|\mathbf{1}_T\|_2
\nonumber\\
&=
\left(
\mathbb{E}_{\zeta_1}^{\pi,\mu_\star^1}
\|m_\star^1\|_2^2
\right)^{1/2}
+
|A|\,|\zeta_1-\zeta_0|
\|\mathbf{1}_T\|_2
\nonumber\\
&\le
\frac{c_0}{16}
\sqrt{\min\{C_T,T\}}
+
|A|c_0
\sqrt{\min\{C_T,T\}}
\nonumber\\
&=
c_0\left(
|A|+\frac{1}{16}
\right)
\sqrt{\min\{C_T,T\}}
\nonumber\\
&\le
\sqrt{\frac{\min\{C_T,T\}}{D}}
\le
\sqrt{\frac{C_T}{D}}.
\label{hatm_feasible}
\end{align}
where we use the definition of $c_0$.
Therefore,
$
D
\mathbb{E}_{\zeta_0}^{\pi,\widehat\mu}
\|\widehat m\|_2^2
\le C_T,
$
and hence
$\widehat\mu\in
\mathcal{S}^{\mu}(C_T;\zeta_0, \pi)$.

Now, we establish the contradiction. From \eqref{eq:follower-square},
\begin{align}
\begin{aligned}
&J_F(\mu_{\star}^0;\zeta_0,\pi)
-J_F(\widehat\mu;\zeta_0,\pi)\\
&\quad=D\left(\mathbb E_{\zeta_0}^{\pi,\mu_\star^0}\|m_{\star}^0\|_2^2-\mathbb E_{\zeta_0}^{\pi,\widehat\mu}\|\widehat m\|_2^2\right)\\
&\qquad\quad+q\left(\mathbb E_{\zeta_0}^{\pi,\mu_\star^0}\|z^0-z^\star(\zeta_0)\mathbf1_T\|_2^2
-\mathbb E_{\zeta_0}^{\pi,\widehat\mu}\|z^{\pi,\zeta_0,\widehat\mu}-z^\star(\zeta_0)\mathbf1_T\|_2^2\right),
\label{eq:followers_objective}
\end{aligned}
\end{align}
where the $T a^2 \sigma_v^2$ terms cancel each other out. We show that the right-hand side exceeds $\eta_T$, contradicting approximate optimality of $\mu_\star^0$. To that end we start with lower bounding
$\mathbb E_{\zeta_0}^{\pi,\mu_\star^0}\|z^0-z^\star(\zeta_0)\mathbf1_T\|_2^2
-\mathbb E_{\zeta_0}^{\pi,\widehat\mu}\|z^{\pi,\zeta_0,\widehat\mu}-z^\star(\zeta_0)\mathbf1_T\|_2^2$,
Since
$
z^0=\zeta_0\mathbf1_T+e^0,
$
we have
\begin{align}
\begin{aligned}
\mathbb E_{\zeta_0}^{\pi,\mu_\star^0}\|z^0-z^\star(\zeta_0)\mathbf1_T\|_2^2
&=T(\zeta_0-z^\star(\zeta_0))^2
+2(\zeta_0-z^\star(\zeta_0))\mathbb E_{\zeta_0}^{\pi,\mu_\star^0}\langle\mathbf1_T,e^0\rangle
+\mathbb E_{\zeta_0}^{\pi,\mu_\star^0}\|e^0\|_2^2\\
&\ge T(\zeta_0-z^\star(\zeta_0))^2
-2|\zeta_0-z^\star(\zeta_0)|\sqrt{T\mathbb E_{\zeta_0}^{\pi,\mu_\star^0}\|e^0\|_2^2}\\
&\ge T(\zeta_0-z^\star(\zeta_0))^2
-\frac{c_0}{8}|\zeta_0-z^\star(\zeta_0)|\sqrt{T\min\{C_T,T\}}.
\label{eq:z0_tracking}
\end{aligned}
\end{align}
where the first inequality follows from Cauchy--Schwarz and the second inequality follows from \eqref{eq:small-errors-revised}. On the other hand, from \eqref{eq:imitation-revised},
$z^{\pi,\zeta_0,\widehat\mu}=\zeta_1\mathbf1_T+e^1$ pathwise under the given coupling, which implies that
$\mathbb E_{\zeta_0}^{\pi,\widehat\mu}\|z^{\pi,\zeta_0,\widehat\mu}-z^\star(\zeta_0)\mathbf1_T\|_2^2=
\mathbb E_{\zeta_1}^{\pi,\mu_\star^1}\|e^1+\zeta_1\mathbf1_T-z^\star(\zeta_0)\mathbf1_T\|_2^2$.
Hence,
\begin{align}
\begin{aligned}
&\mathbb E_{\zeta_0}^{\pi,\widehat\mu}
\|z^{\pi,\zeta_0,\widehat\mu}-z^\star(\zeta_0)\mathbf1_T\|_2^2\\
&\quad =T(\zeta_1-z^\star(\zeta_0))^2
+2(\zeta_1-z^\star(\zeta_0))
\mathbb E_{\zeta_1}^{\pi,\mu_\star^1}\langle\mathbf1_T,e^1\rangle+\mathbb E_{\zeta_1}^{\pi,\mu_\star^1}\|e^1\|_2^2\\
&\quad\le T(\zeta_1-z^\star(\zeta_0))^2
+2|\zeta_1-z^\star(\zeta_0)|
\sqrt{T\mathbb E_{\zeta_1}^{\pi,\mu_\star^1}\|e^1\|_2^2}+\mathbb E_{\zeta_1}^{\pi,\mu_\star^1}\|e^1\|_2^2\\
&\quad\le T(\zeta_1-z^\star(\zeta_0))^2
+\frac{c_0}{8}|\zeta_1-z^\star(\zeta_0)|\sqrt{T\min\{C_T,T\}}+\frac{c_0^2}{256}\min\{C_T,T\}.
\end{aligned}
\label{eq:mhat_tracking}
\end{align}
where similar to \eqref{eq:z0_tracking} we use Cauchy--Schwarz and \eqref{eq:small-errors-revised}. By the definition of $\zeta_1$, it lies between $\zeta_0$ and
$z^\star(\zeta_0)$, and
\[
|\zeta_1-z^\star(\zeta_0)|
=|\zeta_0-z^\star(\zeta_0)|
-c_0\sqrt{\frac{\min\{C_T,T\}}{T}}.
\]
Consequently,
\begin{align}
\begin{aligned}
&T\left[(\zeta_0-z^\star(\zeta_0))^2
-(\zeta_1-z^\star(\zeta_0))^2\right]=2c_0|\zeta_0-z^\star(\zeta_0)|\sqrt{T\min\{C_T,T\}}
-c_0^2\min\{C_T,T\}.
\label{eq:tracking_main_gain}
\end{aligned}
\end{align}

Subtracting \eqref{eq:mhat_tracking} from \eqref{eq:z0_tracking}, and using
$|\zeta_1-z^\star(\zeta_0)|\le|\zeta_0-z^\star(\zeta_0)|$ with \eqref{eq:tracking_main_gain} imply
\begin{align}
\begin{aligned}
&\mathbb E_{\zeta_0}^{\pi,\mu_\star^0}\|z^0-z^\star(\zeta_0)\mathbf1_T\|_2^2
-\mathbb E_{\zeta_0}^{\pi,\widehat\mu}\|z^{\pi,\zeta_0,\widehat\mu}-z^\star(\zeta_0)\mathbf1_T\|_2^2\\
&\quad\ge2c_0|\zeta_0-z^\star(\zeta_0)|\sqrt{T\min\{C_T,T\}}
-c_0^2\min\{C_T,T\}\\
&\qquad-\frac{c_0}{4}|\zeta_0-z^\star(\zeta_0)|\sqrt{T\min\{C_T,T\}}
-\frac{c_0^2}{256}\min\{C_T,T\}\\
&\quad\ge\sqrt{T\min\{C_T,T\}}
\left(\frac{7c_0}{4}|\zeta_0-z^\star(\zeta_0)|
-\frac{257}{256}c_0^2\right).
\label{exiting_eq}
\end{aligned}
\end{align}
where we used $\min\{C_T,T\}\le\sqrt{T\min\{C_T,T\}}$. Since, by definition,
$c_0\le|\zeta_0-z^\star(\zeta_0)|/4$,
we have
\begin{align}
\begin{aligned}
&\frac{7c_0}{4}|\zeta_0-z^\star(\zeta_0)|
-\frac{257}{256}c_0^2\ge\frac{7c_0}{4}|\zeta_0-z^\star(\zeta_0)|
-\frac{257c_0}{1024}|\zeta_0-z^\star(\zeta_0)|
\ge c_0|\zeta_0-z^\star(\zeta_0)|.
\end{aligned}
\end{align}
Hence, \eqref{exiting_eq} implies
\begin{equation}
\begin{aligned}
&\mathbb E_{\zeta_0}^{\pi,\mu_\star^0}\|z^0-z^\star(\zeta_0)\mathbf1_T\|_2^2
-\mathbb E_{\zeta_0}^{\pi,\widehat\mu}\|z^{\pi,\zeta_0,\widehat\mu}-z^\star(\zeta_0)\mathbf1_T\|_2^2 \ge c_0|\zeta_0-z^\star(\zeta_0)|\sqrt{T\min\{C_T,T\}}.
\end{aligned}
\label{eq:tracking_gain_final}
\end{equation}

Returning to \eqref{eq:followers_objective} and using \eqref{eq:tracking_gain_final} yield
\begin{align}
\begin{aligned}
&J_F(\mu_{\star}^0;\zeta_0,\pi)
-J_F(\widehat\mu;\zeta_0,\pi)\\
&\qquad\ge
-D\mathbb E_{\zeta_0}^{\pi,\widehat\mu}\|\widehat m\|_2^2
+q c_0|\zeta_0-z^\star(\zeta_0)|\sqrt{T\min\{C_T,T\}}.
\label{eq:cost_difference}
\end{aligned}
\end{align}
From \eqref{hatm_feasible}, we have
$\bigl(\mathbb E_{\zeta_0}^{\pi,\widehat\mu}\|\widehat m\|_2^2\bigr)^{1/2}
\le c_0(|A|+1/16)\sqrt{\min\{C_T,T\}}$,
and hence
\begin{align}
\begin{aligned}
&J_F(\mu_{\star}^0;\zeta_0,\pi)
-J_F(\widehat\mu;\zeta_0,\pi)\\
&\qquad\ge
-D c_0^2(|A|+1/16)^2\min\{C_T,T\}
+q c_0|\zeta_0-z^\star(\zeta_0)|\sqrt{T\min\{C_T,T\}}\\
&\qquad\ge
c_0\sqrt{T\min\{C_T,T\}}
\left[q|\zeta_0-z^\star(\zeta_0)|
-D c_0(|A|+1/16)^2\right]\\
&\qquad\ge \frac34 q c_0|\zeta_0-z^\star(\zeta_0)|
\sqrt{T\min\{C_T,T\}},\label{eq:epsilon-range}
\end{aligned}
\end{align}
where the final inequality uses
$D c_0(|A|+1/16)^2
\le q|\zeta_0-z^\star(\zeta_0)|/4$.
Choose
\[
\gamma:=\frac38q c_0|\zeta_0-z^\star(\zeta_0)|>0.
\]
By \eqref{eq:epsilon-range}, the cost improvement is at least
$2\eta_T>\eta_T$. This contradicts
\eqref{eq:approx-optimality}, since $\widehat\mu$ is feasible for type
$\zeta_0$. Hence \eqref{eq:small-regret-revised} cannot hold for both
types, and thus
\[
\sup_{i \in \{0,1 \}}R_T^L(\zeta_i,\pi,\mu_\star^i)
\ge\frac{\lambda c_0^2}{256}\min\{C_T,T\}.
\]
Taking the infimum over the follower policies and
using $\{\zeta_0,\zeta_1\}\subseteq\mathcal{Z}$ establishes
the claimed lower bound for every $\pi\in\mathcal{S}^{\pi}$.

\subsection{Proving the $\Omega(\log T)$ lower-bound} 
Set $c_G^{-1}:=8 \sigma_v^{-2}\max\{A^2,1\}>0$. Let $G_T$ be a grid of $\lceil T^{1/4}\rceil$ equally spaced points in $[-\overline{Z}/2,\overline{Z}/2]$, with distance $\Delta:=\overline{Z}/(\lceil T^{1/4}\rceil-1)$ between consecutive points. For each $\zeta\in G_T$, choose an arbitrary policy
$\mu_\zeta\in\mathcal{S}^{\mu}(C_T;\zeta, \pi)$. Define 
\begin{equation}
\begin{aligned}
R:=\max_{\zeta\in G_T}\bigg\{
&\mathbb E_\zeta^{\pi,\mu_\zeta}
\left\|z^{\pi,\zeta,\mu_\zeta}-\zeta\mathbf1_T\right\|_2^2
+\mathbb E_\zeta^{\pi,\mu_\zeta}\|m\|_2^2
\bigg\}.
\end{aligned}
\label{eq:gaussian-lower}
\end{equation}
Let $\widehat Z:=T^{-1}\sum_{t=1}^T z_t$ be the leader's time-averaged estimate, and define the detector $\widehat Z_{\textit{dec}}:=\arg\min_{\zeta'\in G_T}|\widehat Z-\zeta'|$, with ties broken by a fixed rule. Draw $\widetilde Z$ uniformly from $G_T$, independently
of the observation noise, and define the detector's error probability as $P_e:=P(\widehat Z_{\textit{dec}}\neq\widetilde Z)$. Conditional on $\widetilde Z=\zeta$, the interaction
follows the policy pair $(\pi,\mu_\zeta)$. All unqualified probabilities and expectations below
refer to this joint law. We consider two cases to prove that $R\gtrsim\log T$, which, by Lemma~\ref{lem:regret-output}, yields an $\Omega(\log T)$ lower bound on the leader's minimax regret.

\emph{Case 1: $P_e \geq 1/4$.} By the definition of $R$ and Jensen's inequality, $\E[(\widehat Z-\widetilde Z)^2]\le R/T$, where the expectation is taken over the observation noise and the uniformly distributed type on $G_T$. Markov Inequality then yields
\[
  P_e
  \;\le\;
  P\!\left(|\widehat Z-\widetilde Z|\geq \tfrac{\Delta}{2}\right)
  \;\le\;
  \frac{4\E[(\widehat Z-\widetilde Z)^2]}{\Delta^2}
  \;\le\;
  \frac{4R}{T\Delta^2}.
\]
Hence $P_e\ge1/4$ forces
$R\ge T\Delta^2/16\ge \overline{Z}^2\sqrt T/16$.
Since $\sqrt T/\log T\to\infty$, this implies
$R\ge c_G\log T$ for all sufficiently large $T$. 

\emph{Case 2: $P_e<1/4$.}
Since $\widehat{Z}_{\mathrm{dec}}$ is a measurable
function of $(Y_{1:T-1})$, the data processing
inequality and Fano's inequality give
\begin{align}
I(\widetilde Z;Y_{1:T-1})
\ge I(\widetilde Z;\widehat{Z}_{\mathrm{dec}})\ge (1-P_e)\log\lceil T^{1/4}\rceil-\log2> \frac{3}{16}\log T-\log2, \label{lower_boundinbg_with_logT}
\end{align}
where we used $1-P_e>3/4$ and
$\log\lceil T^{1/4}\rceil\ge\frac14\log T$.  The following statement will be proven last. Assume for now that 
\begin{align}
I(\widetilde Z;Y_{1:T-1})
\le
\frac{\max\{A^2,1\}}{\sigma_v^2}R.
\label{final_part_to_be_proven}
\end{align}
Then, combining \eqref{lower_boundinbg_with_logT}
and \eqref{final_part_to_be_proven} gives
\begin{align}
R
\ge
\frac{\sigma_v^2}{\max\{A^2,1\}}
\left(\frac{3}{16}\log T-\log2\right)\nonumber\ge c_G\log T
\end{align}
for all sufficiently large $T$. Thus, in both cases, $R\ge c_G\log T$. Hence, Lemma~\ref{lem:regret-output} implies that
\[
\max_{\zeta\in G_T}
R_T^L(\zeta,\pi,\mu_\zeta)
\ge\lambda c_G\log T.
\]
Since the responses $\mu_\zeta$ were arbitrary, taking the infimum
and using $G_T\subseteq[-\overline{Z},\overline{Z}]$ yields
\[
\sup_{\zeta\in[-\overline{Z},\overline{Z}]}
\inf_{\mu\in\operatorname{BR}^{\eta_T}(\zeta,\pi,C_T)}
R_T^L(\zeta,\pi,\mu)
\ge\lambda c_G\log T.
\]
This establishes the logarithmic lower bound for every
$\pi\in\mathcal{S}^{\pi}$. 

It remains to prove
\eqref{final_part_to_be_proven}. Let $P_{Y_t}$ denote the marginal distribution of $Y_t$,
and let $P_{Y_t\mid\widetilde Z}$ denote its conditional
distribution given $\widetilde Z$. We use the same notation
for conditional distributions given additional variables.
For any fixed reference probability distribution $Q$
on $\mathbb R$ that does not depend on $\widetilde Z$, we have
\[
\E_{\widetilde Z}\!\left[
D_{\mathrm{KL}}(P_{Y_t\mid\widetilde Z}\|Q)
\right]
=
I(\widetilde Z;Y_t)+D_{\mathrm{KL}}(P_{Y_t}\|Q).
\]
By nonnegativity of KL divergence, the left-hand side
therefore upper bounds $I(\widetilde Z;Y_t)$. We apply this identity conditional on $(Y_{1:t-1})$.
Given this history, $z_t$ is fixed, so we choose
$Q_t:=\mathcal N((A+B)z_t,\sigma_v^2)$
as the reference distribution.
This choice depends only on the history and does not
require knowledge of $\widetilde Z$.
Conditional on $(\widetilde Z,Y_{1:t-1})$, the follower's
causal policy also determines $m_t$, and hence
\[
P_{Y_t\mid\widetilde Z,Y_{1:t-1}}
=
\mathcal N(A\widetilde Z+B z_t+m_t,\sigma_v^2).
\]
The KL divergence formula for Gaussian distributions
with equal variance therefore gives
\begin{align}
\begin{aligned}
I(\widetilde Z;Y_t\mid Y_{1:t-1})
&\le
\E\!\left[
D_{\mathrm{KL}}\!\left(
P_{Y_t\mid\widetilde Z,Y_{1:t-1}}\|Q_t
\right)
\right]\\
&=
\frac{1}{2\sigma_v^2}
\E\!\left[
\bigl(A(\widetilde Z-z_t)+m_t\bigr)^2
\right]\\
&\le
\frac{A^2\E[(\widetilde Z-z_t)^2]+\E[m_t^2]}
{\sigma_v^2} \label{mutual_perstep_bound},
\end{aligned}
\end{align}
where the expectations average over the type, and the observation noise. Next, the
chain rule for mutual information and
\eqref{mutual_perstep_bound} give
\begin{align}
I(\widetilde Z;Y_{1:T-1})
&=
\sum_{t=1}^{T-1}
I(\widetilde Z;Y_t\mid Y_{1:t-1})\nonumber\\
&\le
\frac{1}{\sigma_v^2}
\sum_{t=1}^{T-1}
\left(
A^2\E[(\widetilde Z-z_t)^2]+\E[m_t^2]
\right)\nonumber\\
&\le
\frac{\max\{A^2,1\}}{\sigma_v^2}R,
\label{eq:MI}
\end{align}
where the final inequality follows from the definition
of $R$. This proves \eqref{final_part_to_be_proven} and finalizes the proof.

\section{Proof of Theorem \ref{cor:minimax-approx}}

Fix $\zeta\in\mathcal Z$ and
$\mu\in\mathcal{S}^{\mu}(C_T;\zeta,\pi^{\mathrm A})$.
Using $Y_s-Bz_s=A\zeta+m_s+\sigma_v\varepsilon_s$
and the fact that projection onto the convex set $\mathcal Z$
cannot increase the distance between $z_t$ and $\zeta\in\mathcal Z$,
the estimation error $e_t:=z_t-\zeta$ satisfies
\[
|e_t|
\le
\left|
-\frac{\zeta}{t}
+
\frac{1}{At}\sum_{s=1}^{t-1}m_s
+
\frac{\sigma_v}{At}\sum_{s=1}^{t-1}\varepsilon_s
\right|.
\]
Hence, using $(a+b+c)^2\le 3a^2+3b^2+3c^2$ and
$\mathbb E(\sum_{s=1}^{t-1}\varepsilon_s)^2=t-1$,
\begin{align}
\begin{aligned}
\mathbb E_\zeta^{\pi^{\mathrm A},\mu}\|e\|_2^2
&\le
3\zeta^2
\sum_{t=1}^Tt^{-2} + 
\frac{3}{A^2}
\mathbb E_\zeta^{\pi^{\mathrm A},\mu}
\sum_{t=2}^T
\left(
t^{-1}
\sum_{s=1}^{t-1}m_s
\right)^2 + 
\frac{3\sigma_v^2}{A^2}
\sum_{t=2}^T
\frac{t-1}{t^2}.
\end{aligned}
\label{eq:cu-error-bound}
\end{align}
Since $\zeta \in [-\overline{Z},\overline{Z}]$, the first term in
\eqref{eq:cu-error-bound} is uniformly $O(1)$. For the second term, the discrete Hardy inequality gives
\[
\sum_{t=2}^T
\left(
t^{-1}
\sum_{s=1}^{t-1}m_s
\right)^2
\le4\|m\|_2^2
\quad\text{pathwise},
\]
where $m:=[m_1,\ldots,m_T]^\top$ so its expectation is at most
$4\mathbb E_\zeta^{\pi^{\mathrm A},\mu}\|m\|_2^2$. The last term is $\mathcal O(\log T)$ from an integral bound. 
Therefore,
\begin{align}
\mathbb E_\zeta^{\pi^{\mathrm A},\mu}\|e\|_2^2
\lesssim
\log(T)+\mathbb E_\zeta^{\pi^{\mathrm A},\mu}\|m\|_2^2.\label{e_growth}
\end{align}
Finally, Lemma~\ref{lem:regret-output} gives
\begin{align}
R_T^L(\zeta,\pi^{\mathrm A},\mu)
\le\Lambda\left(
\mathbb E_\zeta^{\pi^{\mathrm A},\mu}\|e\|_2^2
+\mathbb E_\zeta^{\pi^{\mathrm A},\mu}\|m\|_2^2
\right)
\lesssim\log(T)+ \mathbb E_\zeta^{\pi^{\mathrm A},\mu}\|m\|_2^2.\label{regret_upper_bound}
\end{align}
Using the expected-budget constraint,
$D\mathbb E_\zeta^{\pi^{\mathrm A},\mu}\|m\|_2^2\le C_T$,
we obtain
\begin{equation}
R_T^L(\zeta,\pi^{\mathrm A},\mu)
\lesssim\log(T)+C_T,
\label{eq:continual-updating-bound}
\end{equation}
uniformly over $\zeta\in\mathcal Z$ and every feasible causal feedback
policy $\mu$. This bound does not require the follower to behave rationally.

We now use the approximate rationality of the follower.
Under $\pi^{\mathrm A}$, consider the zero-manipulation
policy $\mu^0$ where $m_t=0$ for all $t$.
Since $z^\star(\zeta)$ is uniformly bounded over
$\zeta\in\mathcal{Z}$ and $z_t \in \mathcal{Z}$, applying $(a-b)^2\le2a^2+2b^2$
in \eqref{follower_entire_cost} gives
\[
J_F(\mu^0;\zeta,\pi^{\mathrm A})\le K_2T
\]
for some constant $K_2>0$ independent of $T$ and $\zeta$, with no-manipulation. Hence, for every approximately rational response $\mu$ and $m$ produced by such response, \eqref{follower_entire_cost} imply
\[
D\mathbb E_\zeta^{\pi^{\mathrm A},\mu}\|m\|_2^2
\le J_F(\mu;\zeta,\pi^{\mathrm A})
\le K_2T+\eta_T.
\]
which together with the budget constraint implies 
\begin{equation}
D\mathbb E_\zeta^{\pi^{\mathrm A},\mu}\|m\|_2^2
\le\min\{C_T,K_2T+\eta_T\}.
\label{eq:effective-manipulation-budget}
\end{equation}
The regret bound in \eqref{regret_upper_bound} then gives
\[
R_T^L(\zeta,\pi^{\mathrm A},\mu)
\lesssim\log(T)+\min\{C_T,K_2T+\eta_T\}.
\]
For $0 \leq \eta_T\lesssim T$, we have
\[
\begin{aligned}
\log T+\min\{C_T,K_2T+\eta_T\}
\lesssim [C_T]_{\log T}^{T}.
\end{aligned}
\]
Taking the supremum over $\zeta\in\mathcal{Z}$ and
$\mu\in\operatorname{BR}^{\eta_T}
(\zeta,\pi^{\mathrm A},C_T)$ therefore gives the desired result.

\end{document}